\documentclass{PoS}
\usepackage{amsmath,graphicx}

\title{Some Statistics on Women in Lattice QCD}
\ShortTitle{Some Statistics on Women in Lattice QCD}

\author{\speaker{Huey-Wen Lin}\footnote{presenter's email: hwlin@pa.msu.edu}
\\
Department of Physics and Astronomy, Michigan State University, East Lansing, MI 48824 \\
Department of Computational Mathematics,
  Science and Engineering, Michigan State University, East Lansing, MI 48824
}

\abstract{
We present a sampling of analyses concerning the gender ratio of plenary speakers during the years 2000--2016 and make comparisons with other conferences, such as the APS April meeting. We hope this will invite discussion of ideas for how to make our field more accessible to women and minorities. We are preparing for an in-depth survey of the lattice field and welcome any ideas or suggestions. To leave post-conference comments and read about news affecting women in our field, see our Facebook page: https://www.facebook.com/WLQCD
}
\FullConference{34th annual International Symposium on Lattice Field Theory\\
		24-30 July 2016\\
		University of Southampton, UK}

\begin{document}

Have you ever wondered whether female participation is growing in our subfield?
The answer is harder to find out than you might imagine. Even though we just entered the 21st century not so long ago, many of early conference websites no longer exist.
%It is hard to blame anyone, really, who can find a data file that are more than 10 years ago (that would be many laptop generations ago and the battery is probably dead by now; different big data problem there).
Despite the difficulties created by this, we tried to get our hands on as much data as possible and with much help from others in our field, we were able to gather some statistics to answer this question.

\begin{figure}
\begin{center}
\includegraphics[width=0.75\textwidth]{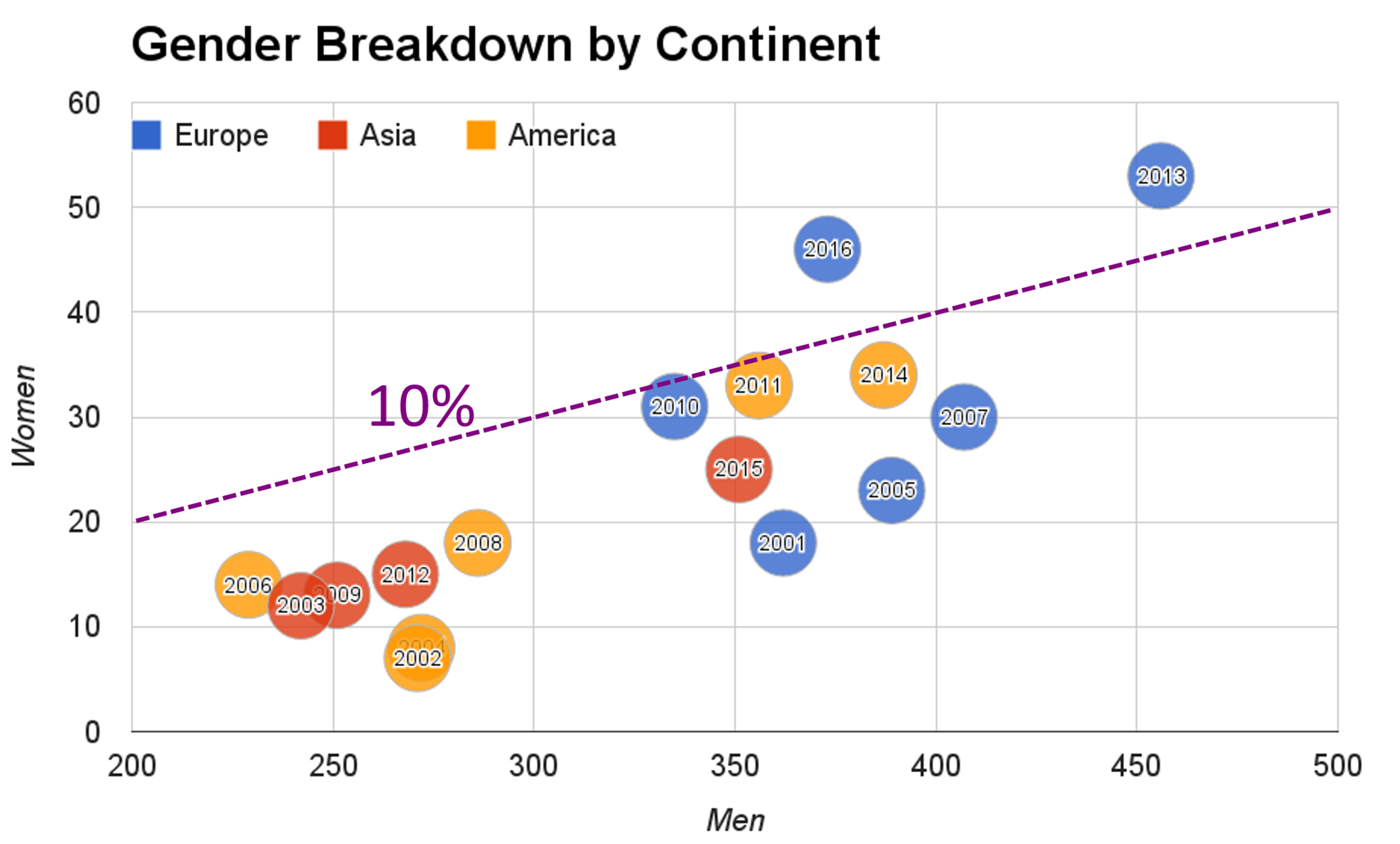}
\end{center}
\caption{The number of female vs male participants; the color indicates the geographic region where the conference was held. blue: Europe, yellow: America, red: Asia/Pacific }
\label{fig:w-vs-m-conf}
\end{figure}

Figure~\ref{fig:w-vs-m-conf} shows the number of women and men participating during the year of the conference marked as a data point. We also color the mark depending on the location of the conference (blue: Europe, yellow: America, red: Asia/Pacific) to more easily discern location-related effects. Overall, the number of women has increased, which is encouraging. The plot shows a line denoting 10\% women, and there is definitely a trend over time increasing above it, in addition to the absolute growth associated with the overall growth of our field. 

Interestingly, but perhaps not surprisingly, the European region seems to have more female participants overall. One possibility is that this is due to the higher number of students in the field. For comparison, it is harder for US institutions to take as many students as Europe, since tuition and benefits costs are much higher; the cost of a student can sometimes be as high as that of a postdoc. However, it is hard to probe this further, since conference registration data do not include details such as the level of the participant (students, postdocs, etc.). Collecting such information might also make it easier to track how many of these students become postdocs and later get positions in academia. This might shed light on the related question of what the rate is of women in our field becoming faculty at universities or tenured researchers at laboratories.  Even this may be just touching the surface; with sufficiently good data, we may be able to come up with many better ways to analyze them. 

\begin{figure}
\begin{center}
\includegraphics[width=0.65\textwidth]{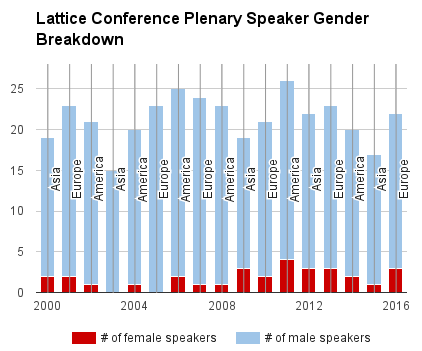} 
\includegraphics[width=0.65\textwidth]{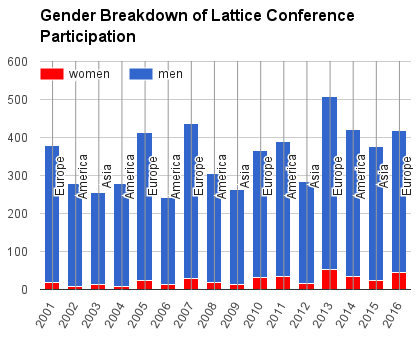} 
\end{center}
\caption{The gender breakdown of plenary speakers (left) and conference participants (right) as a function of year}
\label{fig:gender-breakdown-by-year}
\end{figure}

The next data set we consider, which is somewhat better preserved, is the breakdown of plenary speakers. Plenary talks are often considered an important milestone award for promising postdocs, among those given to senior faculty. 
Are women given these important opportunities for career advancement?
The left-hand side of Fig.~\ref{fig:gender-breakdown-by-year} shows the breakdown of plenary speakers since 2000. Lattice 2011 had the largest number of female speakers in the past 16 years: 4. If one makes a histogram with 5-year bins, then the latest 5 years definitely shows an overall trend of improvement.

Does this correlate with the increasing of the female participants in our field? We compare these on the right-hand side of Fig.~\ref{fig:gender-breakdown-by-year}; unfortunately, the trend is not clear. 

\begin{figure}
\begin{center}
\includegraphics[width=0.45\textwidth]{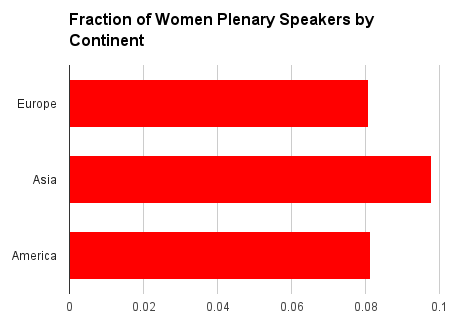} 
\includegraphics[width=0.45\textwidth]{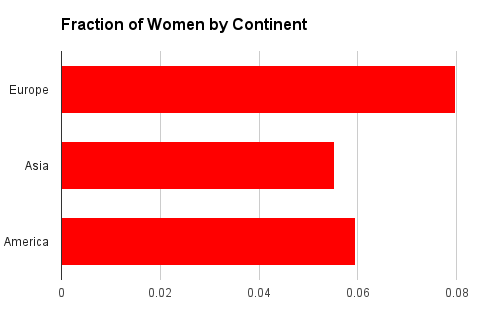} 
\end{center}
\caption{The fraction of women plenary speakers (left) and conference participants (right) as a function of conference location}
\label{fig:gender-breakdown}
\end{figure}

It may be easier to see when we separate the data by continent and take an average of the female fraction over the past 16 years; see Fig.~\ref{fig:gender-breakdown}. It is inconclusive. 

\begin{figure}
\begin{center}
\includegraphics[width=0.75\textwidth]{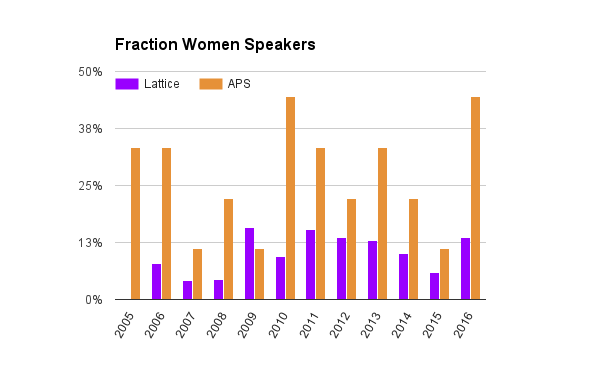} 
\end{center}
\caption{Percentage of female plenary speakers at the lattice conference (purple) and the APS April Meeting (orange) as a function of year}
\label{fig:plenary-comp}
\end{figure}

The highest female-speaker fraction is less than 10\%. Is this good or not? Maybe we can take the APS April plenary data as baseline for comparison; see Fig.~\ref{fig:plenary-comp}. By this metric, we are not doing well at all. Does the Status of Women in Physics Committee provide feedback to the April-meeting's plenary selection that results in the difference? If so, would it be beneficial to have an international lattice women committee to make sure that the yearly selection is being appropriately distributed?
Another possibility is that the difference is due to there being fewer women in theoretical fields than experimental ones. If that were the case, it would have to explain a factor of 2--3 between lattice and the APS. Still, it maybe more straightforward to compare our numbers with another theoretical-physics conference; perhaps an interested reader has further ideas?

\begin{figure}
\begin{center}
\includegraphics[width=0.75\textwidth]{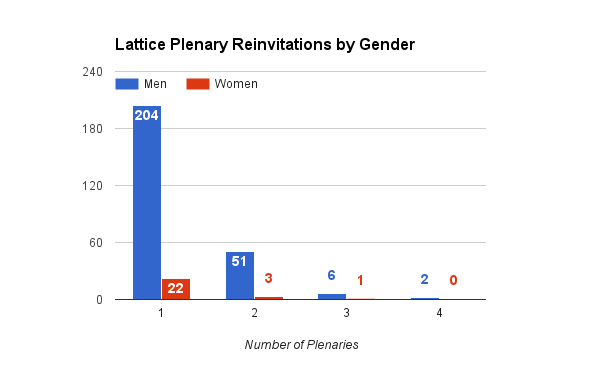} 
\end{center}
\caption{Number of male and female plenary speakers as a function of the number of times those speakers were invited}
\label{fig:reinvitation}
\end{figure}

Having served on the organizing committee before, I have heard the comment that we just do not have enough women in our field.
Can the small number of plenary talks given by women be due to lack of women in our field? Fig.~\ref{fig:reinvitation} shows the gender breakdown for plenary speakers by return rate. 
If we take as a baseline the number of people who gave only 1 plenary talk since 2000, the men-to-women ratio is about a factor of 10. It drops to a factor of 17 for those who return to give a plenary for a second time. There are 8 men who gave plenary talks 3 or more times during this period while only 1 woman in our field has done that. 
So the answer seems to be: yes, we do not have many women in our field to select from to give plenary talks; however, there is definitely room for returning female speakers to improve the ratio imbalance in the current composition.

\begin{figure}
\begin{center}
\includegraphics[width=0.75\textwidth]{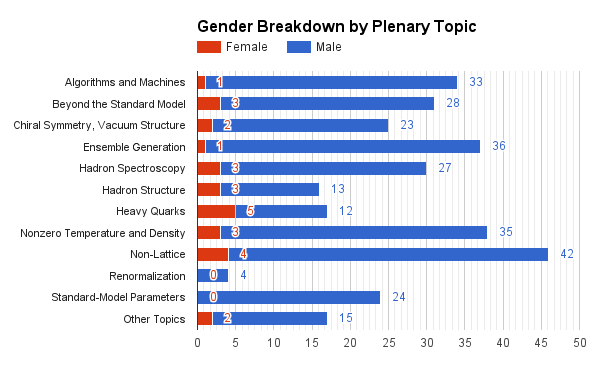} 
\end{center}
\caption{Number of plenaries broken down by gender and topic}
\label{fig:Plenary-by-topic}
\end{figure}

We know there are a number of senior women in flavor physics; do they have more female postdocs or themselves make up the bulk of the female plenary speakers? Fig.~\ref{fig:Plenary-by-topic} breaks down the number of plenary speakers by gender and topic. Note the 5/17 fraction of female speakers in heavy-flavor physics, which greatly exceeds that of any other topic group.
Can this be attributed to a role-model effect? That is, women think they are more likely to succeed in certain subfields since there are already successful women in leadership roles.
A few topics seem to have very low female participation rates; why? 
Are there collaborations where it is harder for women to advance? Do certain topics provide skill sets that women prefer to use to find jobs outside academia? It is hard to conclude without more information.

One thing we know would certainly help with some of these questions is to collect demographics on participants during registration, such as 
gender, position and number of years before or after PhD. We should collect these and make them available online so that these data can be analyzed later.
In particular, we are interested in how we help young people get jobs and whether giving a plenary talk at the lattice conference correlates with landing an academic job? 
Tracing these data may help us get some numbers to improve in the long term. 

We would also like to encourage students and postdocs to come to the 
Women in Lattice QCD luncheon during the lattice conference. It has became a regular event in the conference series on Tuesday during the lunch break. 
It is crucial that Tuesday plenary speakers (and the session chairs) keep to their time limits so that the reserved lunch break is long enough to allow for WLQCD networking and in-depth discussions. 
Everyone can help by avoiding setting collaboration meeting times that overlap such that young women cannot attend WLQCD events even if they want to. 
There are benefits to students and young postdocs:
to network with senior women in our field, 
seek mentorship for career-path planning and
guidance on personal issues, like problems working with male colleagues.

Finally, we would like to once again advertise the Women in Lattice QCD Facebook page: https://www.facebook.com/WLQCD/. It is public so that senior people who do not have a Facebook account can see posts as well.
The Facebook page really needs everyone's help! 
Please send us academic news about women in our field: awards, promotions, workshop organization, big plenary talks, or whatever catches your eye.
We hope the page can also provide a forum for information exchange (Facebook account would be required in this case). 
Please also feel free to leave further suggestions or recommendations for improving it.

%%%%%%%%%%%%%%%%%%%%%%%%%%%%%%%%%%%%%%%%%%%%%%%%%%%%%%%%%%%%%%%%%%%%%%%%%%%%%%%
\vspace{-0.4cm}
\section*{Acknowledgments}
\vspace{-0.3cm}
Those who provided help or comments with the information provided in this study (in chronological order)
Saul~D. Cohen, Andreas Kronfeld, Elizabeth Freeland, Daniel Mohler, Meifeng Lin, Chris Dawson, Shigemi Ohta, Giancarlo Rossi, Doug Toussaint, Derek~B. Leinweber, Vera Guelpers, Etsuko Itou.

%%%%%%%%%%%%%%%%%%%%%%%%%%%%%%%%%%%%%%%%%%%%%%%%%%%%%%%%%%%%%%%%%%%%%%%%%%%%%%%

\end{document}